\documentclass[format=acmsmall, screen=true, review=false]{acmart}
\AtBeginDocument{%
  \providecommand\BibTeX{{%
    \normalfont B\kern-0.5em{\scshape i\kern-0.25em b}\kern-0.8em\TeX}}}
\usepackage{graphics}
\usepackage{multirow}
\usepackage[table,xcdraw]{xcolor}
\usepackage{graphicx}
\usepackage{subcaption}
\usepackage{longtable}
\usepackage{float}

\copyrightyear{2026}
\acmYear{2026}
\acmConference[arXiv Preprint]{arXiv Preprint}{November 5, 2025}{Anywhere, The World}
\acmBooktitle{arXiv Preprint, November 5, 2025, Anywhere, The World}

\begin{document}

\title[Human Resource Management and AI: A Contextual Transparency Database]{Human Resource Management and AI: A Contextual Transparency Database}

\author{Ellen Simpson}
\email{ellen.simpson@virginia.edu}
\orcid{0000-0003-0387-7329}
\affiliation{%
 \institution{University of Virginia}
 \streetaddress{1919 Ivy Road, P.O. Box 400249}
 \city{Charlottesville}
 \state{Virginia}
  \country{USA}
 \postcode{22903}
 }

 \author{Ryan Ermovick}
 \orcid{0009-0008-6864-5637}
 \affiliation{%
 \institution{University of Virginia}
 \streetaddress{1919 Ivy Road, P.O. Box 400249}
 \city{Charlottesville}
 \state{Virginia}
  \country{USA}
 \postcode{22903}
 }

 \author{Mona Sloane}
 \orcid{0000-0003-1049-2267}
 \email{mona.sloane@virginia.edu}
  \affiliation{%
 \institution{University of Virginia}
 \streetaddress{1919 Ivy Road, P.O. Box 400249}
 \city{Charlottesville}
 \state{Virginia}
  \country{USA}
 \postcode{22903}
 }

\renewcommand{\shortauthors}{Simpson, Ermovick, and Sloane}

\begin{abstract}
AI tools are proliferating in human resources management (HRM) and recruiting, helping to mediate access to the labor market. As these systems spread, profession-specific transparency needs emerging from black-boxed systems in HRM move into focus. Prior work often frames transparency technically or abstractly, but we contend AI transparency is a social project shaped by materials, meanings, and competencies of practice. This paper introduces the Talent Acquisition and Recruiting AI (TARAI) Index, situating AI systems within the social practice of recruiting by examining product functionality, claims, assumptions, and AI clarity. Built through an iterative, mixed-methods process, the database demonstrates how transparency emerges: not as a fixed property, but as a dynamic outcome shaped by professional practices, interactions, and competencies. By centering social practice, our work offers a grounded, actionable approach to understanding and articulating AI transparency in HR and provides a blueprint for participatory database design for contextual transparency in professional practice.

\end{abstract}



\maketitle

\section{Introduction}
    Human Resource Management (HRM), the organizational practice of recruiting, supporting, and managing employees \cite{Walker2024}, is experiencing an AI boom. Technologies that support HRM's goals of ensuring an organization meets its operational and business goals (HR Tech) through staffing, supporting, and training the workforce, occupying an estimated global market share of USD 16.43 billion in 2023 \cite{GrandViewResearch2025}, and AI-supported Human Resource Technology products, USD 5.9 billion of that figure in 2023 \cite{MarketUs2024}. At the same time, AI's black box--that is, the inability to reverse-engineer the outputs of an AI system \cite{von2021transparency,pasquale2015black}--is plaguing HRM, leading to biases and inefficiencies \cite{ochmann2021evaluation, Lytton2024AIRecruitingBias,bogen2018help,armstrong2023navigating}. Consequently, there is increased interest and demand for AI transparency--clarity and openness about how an AI model is deployed within an HRM process and how it produces its outcomes so that these can be understood and evaluated by humans \cite{jonker2025transparency,turri2024transparency, sloane2023introducing}. 
    
    Existing human-computer interaction (HCI) research on HRM has focused on the organizational turn in the 1990s \cite{grudin2012moving} and, more recently, questions of human-AI interaction, where researchers set out to explore why HR practitioners resist using AI to evaluate workplace performance \cite{park2021human,park2022designing}. Thus far, however, HCI research has not addressed the issue of \textit{profession-specific AI transparency needs}. This paper fills this gap and is concerned with AI products embedded into HR Tech--a suite of tools, such as platforms, software suites, and automation tools, used by organizations to streamline and optimize the myriad processes of HRM  \cite{RipplingWhatIsHRTech, bogen2018help}.\par

    One area of increased research and regulatory scrutiny around AI interventions is HR Tech products used in a specific area within HRM that is considered high-stakes, as it mediates access to the labor market: talent acquisition and job candidate recruitment (collectively, ``recruiting'') \cite{sloane2023automation,sloane2025systematic}. The profession of recruiting is concerned with sourcing and then vetting potential candidates for fitness for hire into a specific role within an organization. Concerns over the black-boxed nature of the AI used in recruiting have led to increased public scrutiny \cite{Oliver2023, Weed2021, Glasser2024AIHiring,schellmann2024algorithm} and regulatory attention.  For example, AI systems embedded into recruiting have been deemed ``high risk'' for disparate impact by the U.S. Equal Employment Opportunity Commission (EEOC) \cite{Garon2024EEOC}. New York City instituted the Local Law 114 mandating disclosure of the use of automated decision systems [ADS] in employment \cite{NYCLocalLaw144_2021}, and the European Union's AI Act \cite{EUAIAct2024} categorizes AI used in hiring and employment as ``high risk''. \par
    
    The regulatory attention AI in recruiting gets is often oblivious to the social and professional logics of HRM and recruiting specifically, particularly the distributed nature of human-machine decision-making in recruiting and the process of creating meaning across these actors and decision points. Recruiting (and hiring) is rarely just one decision \cite{bogen2018help}. Rather, it is a series of decisions made by the interplay of human- and machine-assessment and decision-making within HRM \cite{bogen2018help}. Regulatory interventions mandated to address AI issues, from the black box to bias, are therefore not grounded in the profession-specific process of meaning-making. We argue that it is particularly relevant for AI transparency, because AI transparency needs are profession-specific \cite{sloane2023introducing}.   
    
    A social theory approach can help lay the groundwork for addressing this problem. We suggest that the social practices \cite{shove2012dynamics,shove2007design} of recruiting (subsequently also: the professional practices of recruiting) can be broken up into three elements that link up and stabilize the practice across institutions: HR Tech products represent the \textbf{material} elements, the physical and digital tools, of recruiting through which the core \textbf{competencies} of the practice of recruiting are enacted and \textbf{meaning} is created and negotiated  \cite{harvey2024cadaver}. By focusing on social practices, we can look beyond a single object, person, or system to more broadly explore the behaviors and practices around the interplay of human and machine-informed decisions or information within a professional context over time \cite{shove2012dynamics,shove2007design,harvey2024cadaver}. We contend that meaning-making and the use of AI in recruiting are co-constructed through interactions with technical systems, peer groups, and the development of competencies around the use of AI broadly within recruiting. \par
    
    We take AI transparency that meaningfully connects to the decisions that have to be made in professional contexts, spreads across the three elements of material, competencies, and meaning, and, therefore, must be grounded in in-depth research of the professional practice---in this case, of recruiting. Yet, much of the scholarship and practice on AI transparency, including explainable AI (XAI) research, frames AI transparency exclusively as a technical project \cite{gunning2021darpa, miller2017explainable}. Here, the black-boxed nature of AI system outputs is addressed primarily by creating a secondary AI model to explain and create model-specific transparency around AI behaviors  \cite{rudin2019stop,martens2025beware} and can potentially mislead users \cite{poursabzi2021manipulating, Wilson2021pymetrics}. \par

    AI transparency matters in professional and social practice, as understanding how AI systems are used in everyday professional use is a key part of meaning-making in a professional context. This approach--called \textit{contextual AI transparency}--mandates empirical engagements with communities of practice, and from that engagement, building on the knowledge of that community to create AI transparency \cite{sloane2023introducing}. In this paper, we work with recruiting practitioners to understand the AI used in recruiting and talent acquisition. Based on the data from these engagements, we focus on contextual AI transparency as \textbf{product functionality, claims, assumptions}, and \textbf{AI clarity}. We create an AI transparency-focused database of HR Tech products that is holistic and both practice-based, situated within HR Tech product information. Our work is part of a growing body of research in HCI that forwards database building as a means of building context-aware AI transparency \cite{feffer2023aiharmsed,walker2024deepfakedb,mcgregor2021AIIncident}. In constructing The Talent Acquisition and Recruiting AI (TARAI) Index\footnote{https://www.tarai.org}, we address the following research questions:

        \begin{itemize}
            \item \textbf{1. What is the recruiting HR Tech space?}
            \item \textbf{2. How is AI deployed within these products?}
            \item \textbf{3. What is being conveyed about AI within these products?}
        \end{itemize}
        
In the sections below, we discuss related work, introduce our database, and then detail the methods by which we built our database. Then, we document our observations and reflections on its construction, focusing on the challenges of creating a database for profession-specific AI transparency. By centering social practice, our work offers a more grounded, actionable approach to understanding and articulating AI transparency in HRM. \par

\section{Related Work}
\subsection{AI in Human Resources in HCI Research} 
  HR Tech are a suite of digital tools, such as platforms, software suites, and automation tools, used by organizations to streamline and optimize the processes of HRM to ensure that companies are working at peak efficiency \cite{RipplingWhatIsHRTech}. HR Tech typically spans all of HRM, from recruiting and hiring, to performance management and workplace culture maintenance, to off-boarding departing employees. Importantly, HR Tech tools typically have common affordances, such as structured data fields for common details (e.g., name, education, experience) that translate between organizations \cite{ajunwa2019platforms}. HR Tech supports the HRM ecosystem \cite{snell2023hrecosystem}, which allows for the development of technologies with narrow uses that are only applicable within specific domains \cite{ajunwa2019platforms}. Due to their specialized nature, HR Tech products typically do not have applications outside of a human resources context. \par
     
     A subset of these specialized technologies are used in job candidate recruitment. The profession of recruiting is concerned with sourcing and then vetting potential candidates for hiring within an organization. HR Tech used in recruiting are technologies that are used to track job candidates while they are being recruited (e.g., Applicant Tracking Systems [ATS]), sourcing and screening technologies, and tools that support interviewing and background checks \cite{bogen2018help,ajunwa2019platforms,sloane2022silicon}. \par

    HCI has explored the topic of HR Tech, focusing on topics such as the design of tools to help lower chances of discrimination \cite{leung2020race}, exploring how generative AI models used in hiring can reflect and enforce stereotypes about gender and racial groups, \cite{Armstrong2024Silicon}, and how gender and race can impact how gig workers are ranked on freelance platforms such as Fiverr \cite{hannak2017bias}. Other relevant research explored the experiences of recruiters \cite{Lashkari2023Recruiterperspectives}, groups of job candidates in hiring while interacting with AI, such as Computer Science students applying for their first jobs\cite{armstrong2023navigating}, or women's experiences with AI candidate ranking and recommendation \cite{chen2018investigating}. Others focused more on the experiences of hiring generally, such as exploring how social class impacts experiences during a job search \cite{chua2021playing}. There is also a significant body of work in HCI-related spaces that explores how to best audit the algorithms involved in hiring \cite{Wilson2021pymetrics, sloane2022silicon}. 

    \begin{table*}[h]
        \centering
        \begin{tabular}{|l|l|l|}
\hline
\textbf{Hiring Stage} & \textbf{Function in Hiring Process} & \textbf{Example HR Tech Product} \\ \hline
\textit{Advertising} & \begin{tabular}[c]{@{}l@{}}Advertises available job, text\\ usually written by a recruiter.\end{tabular} & \begin{tabular}[c]{@{}l@{}}$\bullet$  ZipRecruiter\\ $\bullet$ LinkedIn Jobs\end{tabular} \\ \hline
\textit{Sourcing} & \begin{tabular}[c]{@{}l@{}}Finds candidates to apply to\\ jobs, often focuses on divers-\\ ifying the hiring pool.\end{tabular} & \begin{tabular}[c]{@{}l@{}}$\bullet$ SeekOut\\ $\bullet$  LinkedIn Recruiter\\ $\bullet$ Internal and Past \\ Applicant Pool\end{tabular} \\ \hline
\textit{\begin{tabular}[c]{@{}l@{}}Screening \& \\ Skill Assessment\end{tabular}} & \begin{tabular}[c]{@{}l@{}}Assesses candidates for \\ skills, salary requirements,\\ location, and experience.\end{tabular} & \begin{tabular}[c]{@{}l@{}}$\bullet$  CodeSignal\\ $\bullet$  Internal ATS Chatbots\\ $\bullet$  HackerRank\end{tabular} \\ \hline
\textit{Interviewing} & Interview Candidates. & \begin{tabular}[c]{@{}l@{}}$\bullet$  Hirevue (AI Interviews)\\ $\bullet$  Calendly (Scheduling)\end{tabular} \\ \hline
\textit{\begin{tabular}[c]{@{}l@{}}Selecting \& \\ Background Checks\end{tabular}} & \begin{tabular}[c]{@{}l@{}}Transitions candidate to new\\ hire status, ensures accuracy \\ of candidate credentials.\end{tabular} & \begin{tabular}[c]{@{}l@{}}$\bullet$ HireRight\\ $\bullet$ Vetty\end{tabular} \\ \hline
\end{tabular}
\caption{Hiring stages and their functions, with example technologies.}
\Description{A table showing hiring stages and their functions, with example technologies.}
\label{tab:hiringstages}

        \vspace{-5mm}
    \end{table*}

    Recruiting, part of the ``hiring funnel'' \cite{bogen2018help}, is a series of interrelated processes that characterize important steps along a person's journey from seeing a job advert to onboarding as a new hire. The hiring funnel is usually shepherded by recruiters. Job positions are identified and written up by recruiters, and then \textbf{advertised} on various internal and external job boards and through other channels they deem appropriate and effective. Recruiters will sometimes directly \textbf{source} potential job candidates, drawing from pools of prior applicants and social networking sites. Once potential candidates apply, they are \textbf{screened}, which can involve a conversation with a recruiter, a test to assess technical skills, or possibly an interaction with a chatbot. Once screened, selected candidates will move on to \textbf{interviewing}, which could involve a self-taped interaction with a series of prompts, an interview with an AI avatar, or an interaction with an actual human. Finally, once a candidate is \textbf{selected for hiring}, they may also undergo a \textbf{background check}. Table \ref{tab:hiringstages} shows the hiring funnel and some HR Tech products from each stage. Many recruiters, who are feeling the pressure to integrate AI into their professional practice \cite{sloane2023automation}, do not know how the AI works at various stages of the hiring funnel  \cite{Rocha2025AIJobInterviews,schellmann2024algorithm}. There is a need for stronger AI transparency within HR Tech, but the transparency needs to be situated in the professional practice of recruiting. \par

\subsection{Contextual AI Transparency}

    AI transparency generally focuses on two interrelated concepts: AI interpretability and AI explainability. AI interpretability describes the degree to which a human can understand the decision an AI has made \cite{doshi2017towards}, and AI explainability is more focused on the level of human understanding of the internal functions of an AI model as it makes decisions and is trained \cite{linardatos2020explainable}. In a first step, we define AI transparency as the clarity and openness about how an AI model is deployed within a HRM process and how it produces its outcomes so that these can be understood and evaluated by humans \cite{corbett2023interrogating,turri2024transparency, sloane2023introducing}. In a second step, we stipulate that this clarity and openness ought to derive from the profession-specific decisions that are made within specific contexts that are very similar across organizations. This differs from common definitions of AI transparency, where questions of explainability and interpretability of an AI system are filtered through a technical lens, such as a secondary algorithm to explain the decisions of the first system, and suppose that there is a universal idea of transparency \cite{gunning2021darpa, miller2017explainable, rudin2019stop, martens2025beware, poursabzi2021manipulating, sloane2023introducing}. These approaches are limited, because only through \textit{context-aware} transparency efforts can true shortcomings of AI systems be uncovered and addressed effectively. \par

    We ground this research in the concept of \textbf{contextual AI transparency}, which still emphasizes interpretability and explainability, but contends that AI transparency must be tied to specific social practices and their contexts, such as specific professional decision-making processes into which AI systems get embedded\cite{sloane2023introducing}. What may denote contextual transparency for a HR professional on an AI hiring decision, for example, may be very different than the contextual transparency needed for a doctor reviewing an AI's diagnostic assessment of a patient. Contextual AI transparency contends that it is impossible to achieve a notion of universal transparency, as people come to the table with different culturally situated knowledge \cite{haraway2013situated,sloane2023introducing} and practices.  \par

    Contextual transparency does not dismiss but rather includes the idea of \textit{algorithmic} transparency. This technical `how' addresses the explainability of the algorithm (i.e., how does the AI function, and how interpretable are the AI's findings to a common person). In addition to a focus on the technical, contextual AI transparency also focuses on \textit{interaction} transparency (e.g., the `why'), and \textit{social} transparency, or the social `what' of a technical system as it exists within a particular institutional, legal, or sociocultural context  \cite{haresamudram2023three}. This multi-faceted framework acknowledges that AI transparency is inevitably socially-situated and emerges through regular interaction between humans and technical AI systems \cite{sloane2023introducing,bello2025explainability}. Importantly, this has methodological consequences for research into transparency, as we outline in the next section.

\subsubsection{\textbf{Databases as AI Transparency Tool}}
    A growing body of HCI literature is exploring other means to create AI transparency through the construction of databases. For example, users of these databases can learn about current AI harms with the end goal of preventing future harms, fostering accountability, and supporting ethical behavior from technology companies \cite{rodrigues2023artificial}. One such database, the AI Incident Database, states that it is intended to function as a transparency tool to prevent tech companies from repeating past mistakes through creating a public, collective memory of these incidents \cite{mcgregor2021AIIncident}. This database has shown to be a useful educational tool around topics in AI ethics and topics of AI harms \cite{feffer2023aiharmsed} and has been used across multiple research contexts to either craft intervention \cite{agarwal2024criticalinfra} or to apply the same database-building techniques to different research contexts, such as the use of AI-generated deep fake images for journalists \cite{walker2024deepfakedb}.
    
    Other database explorations are more tailored to certain professional practices, such as Fehr and colleagues' \cite{fehr2024trustworthy} exploration of medical AI tools and the lack of publicly available data on such tools. The authors point to the publicly available databases of these tools, maintained by both the U.S. FDA and the European Commission, in an effort to ensure public awareness of such tools and transparency in how they function. These governmentally-controlled databases are a useful transparency tool, but as Fehr and colleagues \cite{fehr2024trustworthy} observe, these databases are opt-in, meaning that the AI transparency they create is limited to the technology developers who choose to participate in these databases \cite{fehr2024trustworthy}. This contrasts with the databases outlined by Rodrigues and colleagues \cite{rodrigues2023artificial}, including the AI Incident Database \cite{mcgregor2021AIIncident}, which are maintained through submissions by the public. In building these databases \cite {mcgregor2021AIIncident, walker2024deepfakedb}, and then using them as research tools \cite{rodrigues2023artificial,agarwal2024criticalinfra,walker2024deepfakedb}, researchers can start to ask social and professional practice-oriented questions about AI transparency within particular contexts as third parties, rather than public entities or company-grounded or mandated investigations. \par
    
    Database building serves as both a collaborative and in-depth research tool for contextual AI transparency that often situates how it evaluates AI within a particular context. For example, database-oriented AI transparency efforts have been explored in political communication, where researchers documented incidents of deepfakes and less sophisticated `cheapfakes,' exploring the emergent risks of generative AI politics and ensuring transparency for journalists and policy makers \cite{walker2024deepfakedb}. Further, Dao and colleagues maintain a database of ``scary AI'' \cite{Dao2022-AwfulAI}, and the ACLU maintains a database of audits of hiring technologies audited for compliance with New York City Local Law 114 \cite{ACLU_2024}. Within HR Tech, groups such as the American Staffing Association (ASA) have created documentation\footnote{https://americanstaffing.net/asa-staffing-tech-center/essential-elements-staffing-technology/} of the existing ecosystems of HR Tech as they pertain to various aspects of human resources to help recruiters and other practitioners locate available tools within various application areas. There is not, to our knowledge, a comprehensive database of HR Tech tools that is focused on creating contextual transparency around the AI used in HR Tech, particularly with regards to the \textbf{assumptions}, \textbf{claims}, \textbf{functionality}, and \textbf{clarity} embedded into AI for HRM. \par 

\subsection{Assumptions, Claims, Functionality, and Clarity in AI Systems}
\begin{figure}[h]
        \centering
        \includegraphics[width=0.4\linewidth]{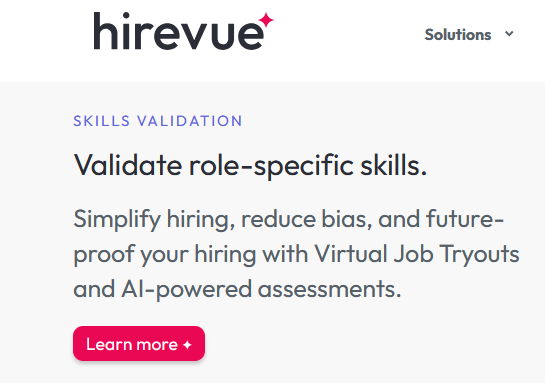}
        \caption{A screenshot of a HR Tech product claim made by the company HireVue.}
        \label{fig:ProductClaim}
        \Description{A screenshot of claims made by the HR Technology company HireVue about their HR tech products. Image follows.}
    \end{figure}

    Embedded into all technologies are opportunities for actions that are afforded by the design of any given product \cite{dourish2001action, davis2020artifacts}. In recruiting, HR tech affordances allow professionals to engage in purposeful activity, such as assessing job candidates, through their use of that product \cite{ratneshwar1999product}. With HR Tech being so niche, the cross-cutting affordances of these tools, namely the way they structure data about people \cite{ajunwa2019platforms}, give rise to a myriad \textbf{functionalities} that still rely on a relatively similar data structure. While two companies may use entirely different HR Tech products, the information stored about people in those products will include data about their tenure, their previous employment, and job performance; it will probably contain a resume with details on a person's education, experience, and career. To this end, many HR Tech companies make specific claims about the functionality of their products as it pertains to HRM, usually outlined in their marketing materials \cite{sloane2022silicon}, such as Figure \ref{fig:ProductClaim}, rather than focus on how their way of structuring data about people is different and/or better. In this example, the company HireVue claims that embedded into the functionality of their product is the ability for a user to validate role-specific skills that a job candidate may possess. \par

    Prior research into HR Tech products has shown that practitioners are interested in actions and outcomes of products -- \textit{functions} \cite{florkowski2018hr, sloane2023automation}. Further, unlike some markets for products (e.g., medical products), which are tightly regulated, HR Tech does not need to provide evidence to support the \textbf{claims }they make about their product's functionality. Instead, HR Tech products will often claim in marketing materials that they have HR best practices embedded into their products, such as being unbiased in hiring, or prioritizing human connection for recruiters, rather than clerical work \cite{wiblen2021digitalised}. This is a flawed approach, according to Wiblen and Marler \cite{wiblen2021digitalised}, who contend that once HR Tech products are implemented and used in alignment with the designer's intentions within an organization, it is the technology and the technology's designers, rather than the recruiter using the technology, that shapes those social processes through which a tool's effectiveness is measured. This research builds on this and other prior work that contends that to understand the social processes of a tool's use within a particular context, one must not look only at the effectiveness and impact of such tools, as is common in HCI research, but also at the claims that these products make \cite{sloane2022silicon}, and the assumptions that underpin the product's functionality \cite{sloane2022silicon}. \par

    Understanding the interplay between claims and product functionality also requires examining the various assumptions that underpin a technology. \textbf{Assumptions} are a set of tacit, background knowledge, and taken-for-granted approaches to any activity or action that are not critically reflected upon \cite{harvey2024cadaver}.  For instance, in Figure \ref{fig:ProductClaim}, the underlying assumption embedded into HireVue's claim about their product's ability to validate job role-specific skills is an assumption that testing and assessment are both an important part of the hiring process, and are an accurate predictor of future job success. Assumptions are a means to understand the underlying social and technical mechanisms that make a product useful to practitioners \cite{sloane2022silicon}. The \textit{normative ideas} in assumptions may clash with professional practice, necessitating the need to identify them and identify where these normative ideas come from \cite{sloane2022silicon}. In the case of Figure \ref{fig:ProductClaim}, there is ample evidence that personality and skill assessments are not universally effective indicators of potential success in a job role \cite{rosse1998impact,kepes2015validity,costa1986personality,furnham2002personality}. To get at these assumptions, one must also understand how a tool or product is used habitually, demonstrating a need to draw not only from textual analysis and ethnographic approaches to explore product marketing materials, but also from conversations with practitioners about their professional practice. \par
  \begin{figure*}
    \centering\includegraphics[width=0.99\linewidth]{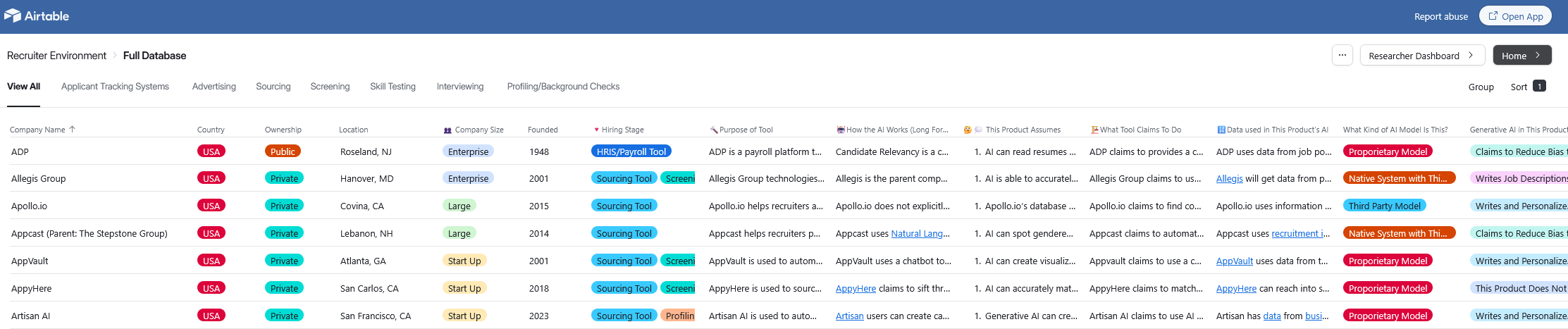}
    \caption{The TARAI Index Recruiter Environment.}
    \label{fig:RecruiterEnv}
    \Description{The TARAI Index Recruiter Environment. Image Follows.}
    \end{figure*}
    
     Consistency and clarity in describing AI systems and how they function are vital for fostering trust in products with embedded AI \cite{park2021human, Kim2025ExplanationsSources}. End-users of these products often express a desire for more context around AI outcomes as a means of creating explainability \cite{kim2023help}. From analysis of a HR Tech's product \textit{functionality}, \textit{claims}, and \textit{assumptions}, a more holistic picture of the \textbf{clarity} of the description of the AI used in an HR Tech product can emerge, adding to its explainability for end-users. For consumers of HR Tech, often these AI are not fully explained without the user purchasing a license to use a HR Tech product. To explore AI clarity as it might be understood as a practitioner investigating HR Tech products, one must first understand how a product works, and be able to discern which of those functions are claims, as opposed to actual functionalities. Through engagement with product marketing websites, one can weigh those functionalities (what does the AI concretely do), claims (what does the website say the AI does), and assumptions (What do those functionalities say about AI, humans, or recruiting), against the marketing materials describing the AI to assess the AI's clarity.  \par

\section{Introducing The TARAI Index}

    The TARAI Index co-constructs contextual AI transparency for the professional practice of recruiting. This overview covers over 100 technology products used in recruiting, with a particular focus on artificial intelligence (AI) systems employed in these products. Information in this database is designed to demonstrate the proliferation of AI in HR Tech for two different communities: practitioners and researchers, catered to in two different database environments.

    In the \textbf{Researcher Environment}, we provide a dashboard offering research perspectives on product claims, assumptions, and AI clarity. The dashboard supports product comparison and sociotechnical analysis within the HR Tech ecosystem and highlights product functionality, claims, assumptions, and AI Clarity. Additionally, these views enable HR professionals to examine product claims about functionality and recruiting.  The Researcher Environment contains four separate research-oriented fields. The environment includes four research fields: AI Functionality and Clarity (Kanban-style), Assumptions and Generative AI (using inclusive tagging to create filtered lists). \par
    
    The \textbf{Recruiter Environment} contains filtering for HR Tech tools as they apply to various stages of the recruiting process (e.g., ATS, screening tools), and offers a separate view to explore product integrations. As Figure \ref{fig:RecruiterEnv} shows, the Recruiter Environment is a big list. The recruiter environment offers a practical overview, allowing recruiters to review product functionality and claims by hiring stage. Each entry on HR Tech products is expandable, providing standardized details on product functionality, claims, assumptions, and AI clarity for comparison..This information is reflected in the Researcher Environment as well, but is tailored to our research questions.. \par
    
    The two environments work in tandem, presenting an understanding of these tools rooted in everyday professional practice as well as the social and technical underpinnings of these technology products, to allow for the exploration of our third line of research inquiry, the enmeshment of HR Tech products, AI, and the social practice of recruiting into each other. \par
    
    The database is using standardized, practice-informed language and is sorted based on the hiring funnel. For recruiters, we provide a comprehensive, cross-hiring funnel function list of products that allow recruiters to sort, search, and manipulate to compare product functionalities, investigate claims and their underlying assumptions, and assess the transparency of various AI integrations within recruiting products, using practitioner-friendly language. For researchers, we provide more tailored information to address the research questions that motivate this project, namely, how HR Tech products are using AI, and how transparent HR tech products are being about that AI. Based on their significance for meaning-making in HR Tech use in recruiting, the core elements of the database are: product functionality, product claims, underlying assumptions, and overall AI clarity.  \par
    
    The TARAI Index draws on over 100 interviews with recruiters and other HR practitioners, using those conversations to develop an understanding of the materials, competencies, and meanings of the professional practice of recruiting. Drawing on this insight, the TARAI Index provides contextual transparency by allowing users to explore available technology options, gain a better understanding of how AI is integrated into various technology tools, and compare available tools. \par

\section{Grounding Our Database in The Professional Practice of Recruiting}    
    In this section, we describe how we used our interview data to ground our research in the professional practice of recruiting, identifying the material objects -- in this case, HR Tech products -- that practitioners commonly used, as well as the competencies of recruiting. In our analysis, we identified several tensions between recruiters and AI, and used these to drive our inquiry into the functions, claims, assumptions, and AI clarity of the HR Tech products we identified. \par

 \begin{figure}
        \centering
        \includegraphics[width=0.90\linewidth]{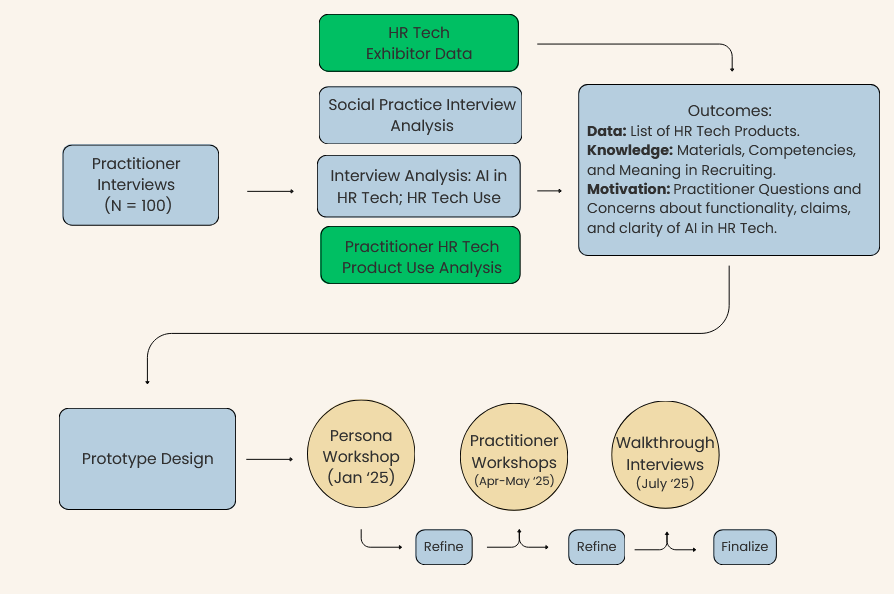}
        \caption{Our research process as a flow chart.}
        \Description{Our research process represented in a flowchart. Image follows.}
        \label{fig:processflow}
    \end{figure}

\subsection{Interviews: Grounding in Social Practice}
    We conducted over 100 semi-structured interviews with HR practitioners, recruiters, and HR Tech developers (2021–2025), across two institutions with IRB approval. Participants were recruited via professional networks, direct solicitation, and social media (LinkedIn).. Participants were at various career stages, and worked in the U.S., Canada, Australia, India, and England. We did not collect demographic information about the practitioners beyond these details. Interviews focused on work practices, HR Tech products used, and perspectives on recruiting.  These interviews were designed to help us understand the social practices of recruiting. Participants mentioned competencies (e.g., being able to read people), materials (e.g., what ATS they used, or have used in the past), and discussed how they negotiated meaning in a changing HR Tech landscape. Interviews (30–90 minutes) were conducted remotely via Zoom, recorded and transcribed with consent. \par

\subsection{Analysis: Understanding Practitioner Transparency Needs}
    To prioritize information for the HR Tech database, we independently analyzed four years of interview data (2021–2025) from HR professionals. Addressing our research questions,  our coding focused on HR tech use in recruiting and AI transparency needs, with our  analysis focused on AI integration into HR tech products. Several themes emerged from this analysis that directly informed the our database framework and design:
    \begin{itemize}
    
        \item \textbf{AI deployed in HR tech products does not map onto the information needs of recruiters}---such as P76, a technical recruiter, who commented: \textit{``None of the AI tools that all these platforms have been releasing have come anywhere close to being as useful [in sourcing] as just basic Boolean operators''} and P99, a staffing manager, pointed out the problem with AI-powered search on tools like LinkedIn: \textit{``LinkedIn Recruiter uses a lot of AI, I can't stand their AI because it limits the amount of information I can find. [...] I think when you just have these standard AI searches, it's really going to limit things for recruiters.''} \par        
        
        \item \textbf{Questions persisted about how the AI that they were using worked and how it reached the conclusions that it did}---P73, an independent recruiter, commented on how recruiters sometimes used AI candidate ranking as an \textit{``an arbitrary sorting mechanism [of candidates].''} P90, a sales director for his staffing agency, pointed out some flaws in ranking systems, explaining: \begin{quote}\textit{``I don't like them, the ranking tools are going to make those rankings based purely on the resume. And resumes never encapsulate the person. They're looking for titles. They're looking for progression. They're looking for skills. They're looking for keywords. So they're looking at the job description that I wrote, comparing that, looking for keywords, and if you have more keywords than somebody else, you're ranked higher, even though all you did was stuff keywords in there, and they didn't.''}\end{quote}\par
        
        \item \textbf{HR Tech products do not allow for nuance in meaning-making}---P80, a senior recruiter, explained his frustrations with the ranking systems used at his job as they lacked nuance: \textit{``Technically, our ATS four options: excellent, good, bad, poor answer and at the end of the interview, you hit score interview, and it would use those answers to create a score card. I did not like that system, because who am I to determine if that's a great answer or a good answer? Like, they answered the question.''}  \par
        \end{itemize}
        
    Cutting across these themes is a threat to professional discretion in recruiting \cite{sloane2023automation}. AI-powered candidate search or candidate ranking is obscuring human-powered discretion in favor of AI-driven hiring decisions, according to the practitioners we spoke to, representing a loss of professional discretion as AI is integrated into HR Tech products. Participants revealed tensions between what a product \textbf{claimed} it could do, and what its \textbf{functionality} was in practice, and they further discussed their concerns about the opacity - and therefore clarity -  of the AI embedded into the HR Tech products they were using.  Opacity combined with automation threatens discretion, as the recruiters we spoke to are used to leveraging tools to expand, not limit, their discretion; much of what recruiters do in their professional practice, for better or for worse, is based on gut feelings. The statements quoted above characterize the themes of many conversations we had with practitioners, and direct our attention back to a need for contextual transparency for recruiters and HR practitioners. \par   
 
\subsection{Secondary Analysis: Identifying A Data Sample of HR Tech Products}
\begin{figure}
        \centering
        \includegraphics[width=0.8\linewidth]{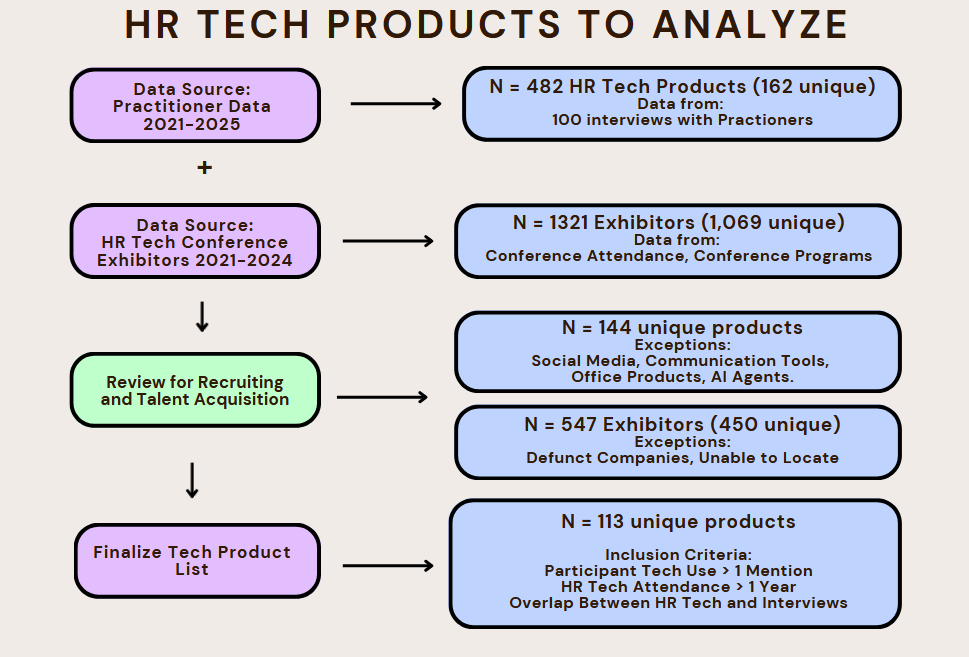}
        \caption{Refining our corpus of HR Tech Products.}
        \Description{Refining our corpus of HR Tech Products. Image Follows.}
        \label{fig:RefiningCorpus}
        \vspace{-5mm}
    \end{figure}

    To identify HR Tech products used in recruiting, we documented HR Tech products mentioned in interviews, and ranked them by the frequency that they were mentioned in the interviews. Second,We gathered exhibitor lists from the annual Human Resources Technologies Conference (Las Vegas, NV, U.S. 2021–2024)\footnote{The Human Resources Technologies Conference takes place annually in September, so we have not included 2025 attendance data.}. We selected the Human Resources Technologies Conference because it is the professional conference that has the closest affiliation with the HR Tech industry, which also includes a large exhibition of HR Tech products. After screening the final list of 1,069 exhibitors for product application in recruiting, we found 450 unique products and compared these with products mentioned in interviews. The list of products mentioned in interviews and the list resulting from the conference exhibitor search were then compared, resulting in our final corpus of product data, with inclusion criteria coming from overlap, multiple interview mentions, or multiple conference appearances. This resulted in 113 unique products. Figures \ref{fig:RefiningCorpus} and \ref{fig:processflow} detail these steps.\par

\section{Building the HR Technologies Database: Sourcing Data and Design}
    In this section, we describe our process for building our prototype database. We describe our data collection method and where we sourced data about HR Tech Companies. We then introduce our considerations when building our prototype database.

\subsection{Text Analysis of Product Marketing Materials}
    To assess AI transparency in 113 HR Tech products, we sourced `claims' data directly from product websites, not external sources, treating them like primary sources. We chose this approach as product websites are often the only information source for practitioners, and therefore should have explainable information.  Reviewing product documentation is an established HCI research method \cite{scheuerman2020genderfacialanalysis, Hedegaard2013ExtractingUX}. If websites products did not contain ``demographic'' information about HR Tech products (e.g., company size, physical address), we used Pitchbook and Crunchbase as these databases are well-established and used in HRM and business schools. We also tracked mergers and acquisitions, noting from the interviews that some participants still used legacy products (e.g., using Taleo as opposed to Oracle's other HRIS products). \par

    The research team reviewed company marketing materials using manual searches on Google using `site:' and Chat GPT-3 to search within specific web domains. We did this for cross-verification, and validated search results against cited text for accuracy. from here, we entered the data manually after comparing our findings and resolving discrepancies with further research as a group. The team met weekly to refine our database schema, design and display conventions, and refine our codebook as new findings emerged. \par

    \begin{figure}[h]
        \centering
        \includegraphics[width=0.99\linewidth]{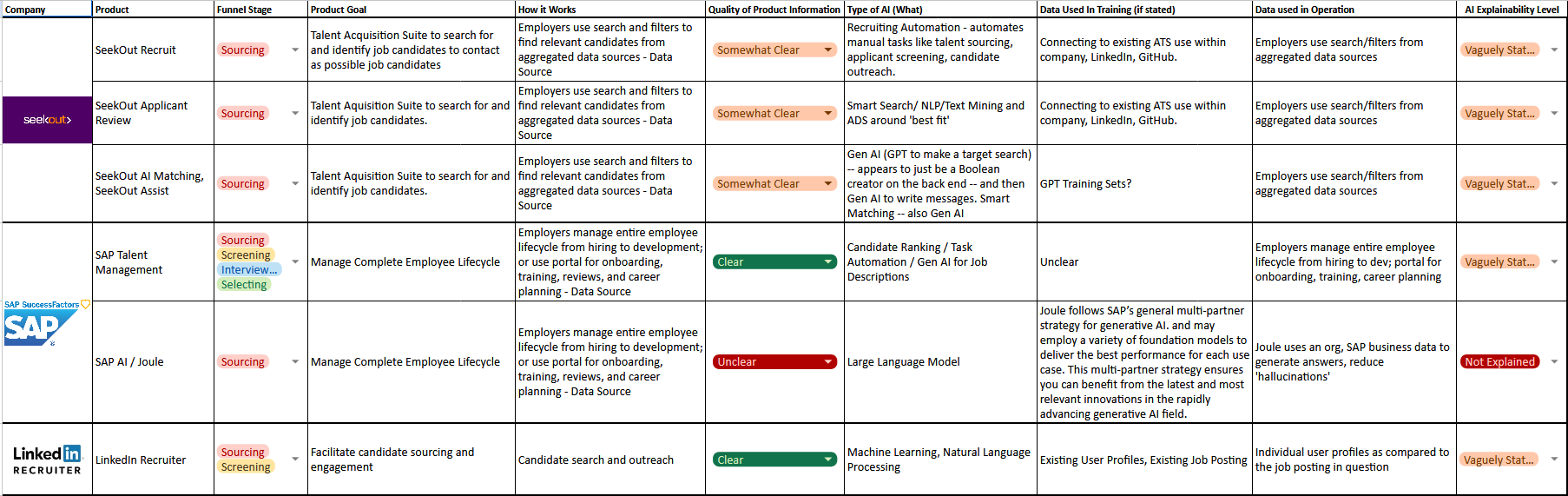}
        \caption{The initial prototype for the HR Tech Database.}
        \label{fig:Database V1}
        \Description{The initial prototype for the HR Tech Database. Image Follows.}
        \vspace{-5mm}
    \end{figure}

    \subsection{Building A Prototype}
    We aimed to make the information in our database accessible by aligning it with recruiters’ professional practices. The practitioners we spoke to described challenges with HR Tech products throughout the hiring process (e.g., P80's issues with his ATS interview evaluation systems), which highlights a need for better tools. We integrated key terms used by recruiters (e.g., `hiring funnel') into the database, as practice-specific language is a `competence' we obtained from interviews \cite{shove2012dynamics}. We determined that sorting HR Tech products by hiring stage and tagging them for easy search based on the hiring funnel would maximize searchability. Additionally, we translated product marketing descriptions into recruiter-friendly language, identified through our interviews, such as recruiter acronyms and lingo like ATS or HRIS. \par
    
     We reviewed product websites to identify supported hiring stages and selected 10 HR Tech products for initial analysis, ensuring that we covered all hiring funnel stages. During our investigation, we found more evidence that AI intransparency is not just a technical problem, but rather one that extends into the superficial product descriptions we encountered that could potentially lead to customer deception. P78, a former corporate recruiter, described frustration with an AI-driven HR Tech product he used in a previous role: 
    
        \begin{quote}
        \textit{``I don't want to say that [anonymous HR Tech product] were purposely obfuscating what they were and weren't doing, but I think a lot of what is going on behind the closed doors with these companies is they can't [say what they're doing]. [...] The biggest frustration I think I have is that there's no education around it. There isn't any awareness. And so if I'm a recruiter using this tool, I have no calibration of whether it's working the way it should, and I potentially cause harm somewhere, right? Like, there isn't any [transparency].''}
        \end{quote}

    P78 explained that he did not think the tools he was using could do everything they claimed they could do, which was why their websites weren't transparent. Other participants identified similar frustrations, with P29, a technical recruiter, explaining, 

        \begin{quote}
        \textit{``I think that is always a question on how the candidates are being ranked [within an ATS], because some candidates rank higher and do not fit for your role. And you wonder why this person popped up? And you typically don't have answers.''}
        \end{quote}
    
    These concerns led us to investigate the initial ten HR tech products for how they talked about AI. We observed inconsistencies here as well, where sometimes \textbf{what an AI was} was explained (e.g., ``LLM'' or ``ChatGPT wrapper')' but its \textbf{function} and use \textbf{context} were often unclear. We found that company disclosures were often the sole source of information on these AI models and their function. When this information was available, we documented AI type, how the AI worked (its function and context of use), and any information on training data for that AI we could find.\par

    In researching these details about how AI functioned, and in gathering basic information about the companies that produce HR Tech products (e.g., company age, location, and number of employees), our work provided more evidence that AI intransparency extended beyond the technical black box into other spaces, such as superficial product descriptions. P100, a recruiter, dubbed many of these products `vaporware,' explaining:

    \begin{quote}
\textit{        ``When companies say we're the next AI hiring platform and [the platform] is just a landing page, [or] there's no software that exists or there's some flashy demo, and it's just mostly like UI mock-ups. And [if] the tool [has] actually been deployed, the data that it's ingesting is irrelevant. You see this all the time with a lot of these companies who will try to sell, like, the world's largest database. And, well, where is this data coming from? And that's not described at all.''}
    \end{quote}
   
     P100's determination that many of these HR Tech products were `vaporware' with unclear data sources illustrates this finding well. He later pointed to frustrations in his own research into HR Tech products, noting that it was challenging to understand even basic information about how these products functioned without having used them himself. \textit{``I certainly am thinking about going down that route where I am engaging actively with these CEOs and asking, can you show me a peek behind the curtain?''} He said when describing how he wanted to get a better sense of which of these products would be beneficial for recruiters to use. \par

    Based on the opaqueness we observed in product functionality, AI clarity, and claims, we added details on AI functionality for each product. This included information about whether a product ranked candidates with AI, if they offered process automation, and if there was integrated generative AI. We developed a tagging system to capture increased generative AI integration in HR Tech, as it was top-of-mind for many interviewees. This includes broad product functionality details such as: ``uses Generative AI to write candidate correspondence'' and ``uses process automation to schedule interviews.'' We developed similar tags that specifically covered generative AI features. We also tagged and summarized product claims and embedded assumptions in long form and in tags. \par

    \begin{figure}
        \centering
        \includegraphics[width=0.99\linewidth]{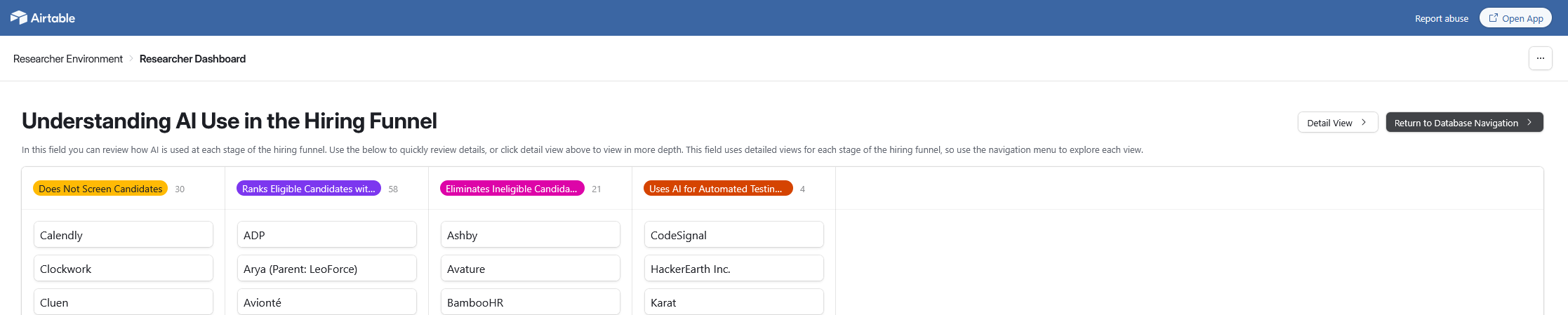}
        \caption{The Researcher Environment; showing the AI Use in the hiring funnel Dataset.}
        \label{fig:ResearcherEnv}
        \Description{A screenshot of the Researcher Environment; showing the AI Use in the Hiring Funnel Dataset. Image Follows}
    \end{figure}

\section{Evaluation Methods}

    We conducted three iterative design workshops to refine the database, following standard design methods \cite{OlearyIterativeDesign2018}. Genuine participation in AI design is not considered possible \cite{sloane2022participation,birhane2022power,delgado2023participatory,madaio2022assessing}, because AI systems design is an exclusionary process. By using iterative, participatory methods, we enabled practitioner input for an AI transparency database grounded in professional practice. Our design process began with stakeholder input, followed by prototyping and rounds of feedback until design goals were met \cite{cross2023design, OlearyIterativeDesign2018}. For this research, we treated our prior interview research as our initial stakeholder meetings; and used the emergent questions that HR practitioners to create an initial prototype. This initial prototype was presented at a public symposium connected to the funding source of this research, where  we asked attendees to take on specific personas that use and assess HR Tech products. Personas are a common way to organize information about users and conduct scenario-based research in HCI \cite{salminen2022usecasespersonas}, and they provide common ground for diverse groups \cite{aljohani2015personas}. The next workshops were held as groups or 1:1 with HR Practitioners at various stages of expertise. The final workshops were more formal technology walkthroughs with HR executives who often make product recommendations for license purchase. We briefly outline the basic procedure of each below. \par 

    \subsection{Initial Workshop: Personas} This workshop took place in January 2025 at a symposium at our home institution, with 50-60 people in attendance. Participants were data science faculty, business school faculty, graduate students, and interested members of the local community. While there was not a specific participant count, as people came and went from the talks, approximately 50-60 people were present, and the workshop lasted 50 minutes. Participants received study information and consent, and reviewed preliminary findings. Due to the wide range of experience among the attendees, we created personas. Personas are useful in design-based research as their introduction into early-stage research introduces hypothetical end-users to drive research through immersion into the hypothetical end-users’ needs \cite{aljohani2015personas,salminen2022usecasespersonas}. For this evaluation, three personas were created: recruiter, HR procurement, and AI auditor (see Appendix A for full details). \par

    Each persona group was led by a member of the research team, assisted by a note taker. Groups assessed the prototype database and provided feedback. Further, participants were free to write on their copies of the database, which we digitized at the end of the workshop.   \par

    \subsection{Second Workshop: Practitioner Workshops}
    After reviewing data from the persona workshop, We revised and expanded the database in Airtable, a database program, increasing the HR Tech products detailed in it from five to thirty based on workshop feedback input. These products were picked from the list of HR Tech tools that practitioners discussed the most, so as to ensure familiarity with the HR Tech products displayed. We developed practitioner-focused navigation interfaces in Airtable, taking advantage of it's design features to make it easier for practitioners to explore the database. \par
    
    The practitioner workshops were held between April and May 2025, with practitipants recruited on social networks such as LinkedIn, and former interviewees who had expressed interest in staying involved with the project. Table \ref{tab:Workshop2Participants} in Appendix A shows the professional role for each participant and their associated workshop session. Eight people signed up to attend the first two workshops, but only one (Workshop \#2) was a collective evaluation among four practitioners. The other evaluations sessions were done one-on-one with the researchers, who walked the participants through how they'd navigate the database, asked questions about what they were seeing, and explored the information the research team had gathered. Participants requested more exploration freedom and less academic language.\par

\subsection{Evaluation: Walkthrough Interviews}
    For the third database version, we aimed to understand professional use and impact of a database like this by HRM professionals. We used an approach based on the walkthrough method \cite{light2018walkthrough} to have participants walkthrough the database and assess its usability. We recruited five experts using UserInterviews.com, and compensated them at a rate of \$60.00 USD/hour (Table \ref{tab:Workshop3Participants} in Appendix A has their demographics). In addition to these five participants, two prior workshop participants and one former interviewee also contributed feedback.. \par

\section{Discussion: Building the HR Technologies Database}
    In this section, we discuss our findings on building the TARAI Index as a method for creating practice-centric AI transparency. In our observations, we found that product functionality was often ambiguously described, creating challenges for AI transparency, which we documented in the database. Additionally, we discuss hour observations on how product claims and assumptions clustered around broader themes.  We include participant feedback on how we represent of claims and assumptions in the database, and discuss describe our assessment of AI clarity and discuss limitations of our approach to creating contextual AI transparency in a database.

\subsection{Characterizing Product Functionality}

    To define product criteria, we first examined product websites and promotional materials to to \textbf{characterize AI types}, such as generative AI or large language models. This process revealed which HR Tech products have generative AI or process automation functionality. These were important functionality indicators, helping to group similar HR Tech products. While making this determination, we also were able to cut through some of the \textit{`junk jargon'} (PW1-1), to focus on actual functionality. We had to be careful, though, as PW3-1 also pointed out that sometimes, \textit{``[T]he feature is bullshit, it doesn't even exist, or it barely exists...''}. We ensured that functionality fields documented only what marketing materials stated. We created tags for common AI and generative AI functionalities. While generative AI functionality was embedded into the broader AI functionality description, generative AI also received separate tags due to practitioner interest during our interviews. \par

    During the practitioner workshops, practitioners highlighted their need to compare products, often due to resource constraints. Comparing products and their integrations was also discussed during the our walkthrough interviews, and was a desire mentioned during our interviews. Walkthrough participants found the database provided useful grounding for newcomers in HRM or recruiting. \par

\subsubsection{\textbf{Tagging and Sorting Stated Functionalities into Clusters}}
\begin{figure}
        \centering
        \includegraphics[width=0.5\linewidth]{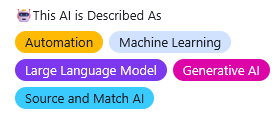}
        \caption{A screenshot of the TARAI Index showing the various ways the company Findem described their AI.}
        \Description{A screenshot of the TARAI Index showing the various ways the company Findem described their AI.}
        \label{fig:Findem}
    \end{figure}

    Our articulation of product functionalities evolved during evaluation sessions. PW2-2, a technical recruiter, mentioned wanting \textit{``just [a] feature list [as] a way to start''}, similar to feedback from the first workshop, where all three persona groups gave feedback to us that the database should provide more detail about product functionalities. We developed inclusive and exclusive tag-based sorting systems for these products. These tags detailed (1) what stages of the hiring funnel these tools supported, (2) generative AI functionality, and (3) overall product functionality, including AI functionality. We also documented the language that companies used to describe any AI in their products. Tagging revealed frequent vague or potentially misleading AI descriptions, often presented next to  clear definitions of other product functionality. For example, Findem describes its AI as ‘Generative AI’ (clear) and ‘source and match AI’ (less clear): what is being sourced? And what is it being matched to? Figure \ref{fig:Findem} shows the various ways Findem described their AI as reflected in the TARAI Index. \par

\begin{figure}
        \centering
        \includegraphics[width=0.98\linewidth]{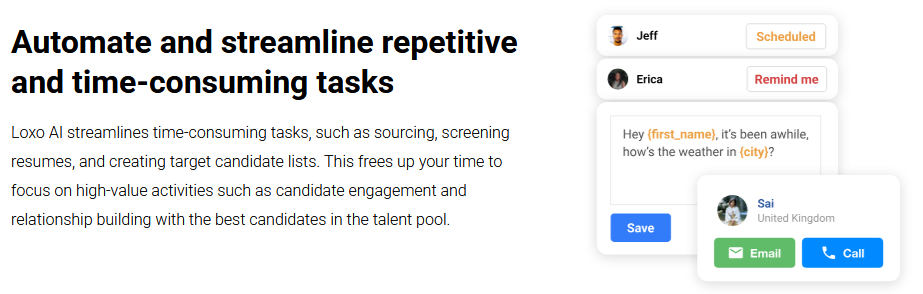}
        \caption{A screenshot of the HR Tech Product, Luxo's website, describing how their product uses process automation.}
        \Description{A screenshot of the HR Tech Product, Luxo's website, describing how their product uses process automation. Image follows.}
        \label{fig:luxostreamline}
    \end{figure}

    These tag clusters helped us find common threads between products and informed our analysis of how HR Tech products process candidate data. Language describing product functionality varied based on what was being discussed. For example, Loxo, an ATS show in \ref{fig:luxostreamline}, uses clear language for automation, conveying how the process works, identifying hiring funnel stages like candidate sourcing and screening. 

\begin{figure}[h]
        \centering
        \includegraphics[width=0.4\linewidth]{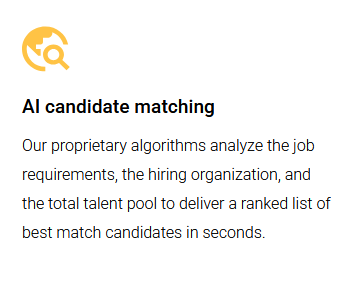}
        \caption{A screenshot of the HR Tech Product, Luxo's website, describing their product's AI.}
        \Description{A screenshot of the HR Tech Product, Luxo's website, describing their product's AI. Image follows}
        \label{fig:loxoAI}
    \end{figure}

    However, Loxo’s description of AI is opaque, as it conflates AI functionality with automation. Figure \ref{fig:loxoAI} shows how a product function, such as using AI to rank job candidates based on job criteria, is clearly described, but couples that with vague language to describe the AI by using `proprietary algorithms.' While both of these descriptions promote themselves as time-saving measures, automation is expressed more clearly via examples, but AI description lack supporting detail. We frequent unevenness in how clearly product functionality was defined, especially around AI evaluating, assessing, or processing candidate data.\par

\subsubsection{\textbf{Deciphering How HR Tech AI Evaluates, Assesses, and Interacts with Candidates}}
    The final functionality field assessed how AI operated within HR Tech products. Unlike our general functionality tagging, we used closed tags to document AI functions. We were concerned and wanted to document AI's impact on the candidate's actual chances of advancing in the hiring process. In our review, we found that there were three broad clusters of AI intervention on candidate hiring, as shown in Table \ref{tab:AIInterventions}.

    \begin{table*}
       \centering
\begin{tabular}{|l|l|l|}
\hline
\multicolumn{1}{|c|}{\textbf{AI Intervention}} & \multicolumn{1}{c|}{\textbf{Explanation}} & \multicolumn{1}{c|}{\textbf{Example Tech}} \\ \hline
\textit{\begin{tabular}[c]{@{}l@{}}Uses AI to Eliminate \\ Ineligible Candidates\end{tabular}} & \begin{tabular}[c]{@{}l@{}}This included, broadly, process automation processes \\ such as knock-out questions candidates completed \\ while submitting an application. It also included the \\ interventions of chatbots, programmed to gather basic \\ candidate information such as willingness to relocate \\ or years of experience and compare it to pre-\\determined eligibility criteria.\end{tabular} & \begin{tabular}[c]{@{}l@{}}HireVue\\ Taleo\\ Jobvite\end{tabular} \\ \hline
\textit{\begin{tabular}[c]{@{}l@{}}Uses AI to Assess \\ Candidate Skill\end{tabular}} & \begin{tabular}[c]{@{}l@{}}This category was primarily concerned with tools \\ designed to assess a job candidate's skill, particularly \\ technical skills (e.g., programming, problem solving) \\ These tools assigned candidates scores, but the \\ scores were not used to advance or reject a job \\ candidate.\end{tabular} & \begin{tabular}[c]{@{}l@{}}CodeSignal\\ HackerEarth\\ Karat\end{tabular} \\ \hline
\textit{\begin{tabular}[c]{@{}l@{}}Ranks Eligible \\ Candidates with AI\end{tabular}} & \begin{tabular}[c]{@{}l@{}}In this category, we identified products that reported \\ an embedded functionality that \textit{ranked} candidate in \\ some way. Sometimes, these candidates had already\\been subject to an elimination around with a chatbot \\or automated process. This included terms  meant to \\ convey that an AI was processing available infor-\\mation about a job candidate and assigning some kind \\ of score in a context that was not focused on \\ evaluating technical or problem-solving skill.\end{tabular} & \begin{tabular}[c]{@{}l@{}}Gem\\ iCIMS\end{tabular} \\ \hline
\textit{No AI} & \begin{tabular}[c]{@{}l@{}}This was a catch-all category for products whose AI \\ did not impact a candidate's ability to advance in the \\ hiring process, or products that did not use AI.\end{tabular} & \begin{tabular}[c]{@{}l@{}}Vetty\\ WorkReels\end{tabular} \\ \hline
\end{tabular}
\caption{Examples of AI Intervention in HR Tech.}
\label{tab:AIInterventions}
\Description{Examples of AI Intervention in HR Tech, table follows.}

    \end{table*}

    Analyzing how an AI functioned in this context revealed the opaque language in marketing for high-risk AI systems. Process automation and AI ranking alike were often collapsed into the nebulous category of `AI.' We observed this consistently, such as the example of Loxo, increasing risk for buyers and practitioners. In using imprecise language, HR Tech products are contributing to AI Transparency issues. Our tagging system product and AI functionalities, including how AI interacts with candidates, were viewed positively. W3-3, explained: \textit{`I think when companies buy tools, especially if [the recruiting team] isn't involved in that decision making process, we don't know the full [product] functionality, because nobody tells us, and then we ask questions, but then it becomes an add on and a cost'} if the functionality she needed wasn't in the purchased product. PW3-1, while discussing our second prototype's initial focus on applicant tracking systems, added, \textit{`Individual recruiters are still looking for functionality [...] one-off narrow, niche products.'} In both of these instances, the TARAI Index was useful in cutting through the opaque marketing language of these HR Tech products, which is often repeated and validated by sales people who, as Pw3-1 explained, find it easy to sell a product that `\textit{had no bugs [and] did everything [a company] asked, and was going to make [that company] completely happy}'---even if that product did not actually exist. Treating AI as a marketing term obscures its functions within a product, reducing transparency. The average user sees automated processes, such as knock-out questions commonly embedded into job application forms that verify licenses, visa status, or sometimes even physical location, as the same AI that is sorting, classifying, and ranking job candidates based on their job materials. This led us to dig deeper into the claims a product was making about what it could do, and what assumptions may have informed that claim.\par    
    \begin{figure*}
\centering
\begin{minipage}{.5\textwidth}
  \centering
  \includegraphics[width=.99\linewidth]{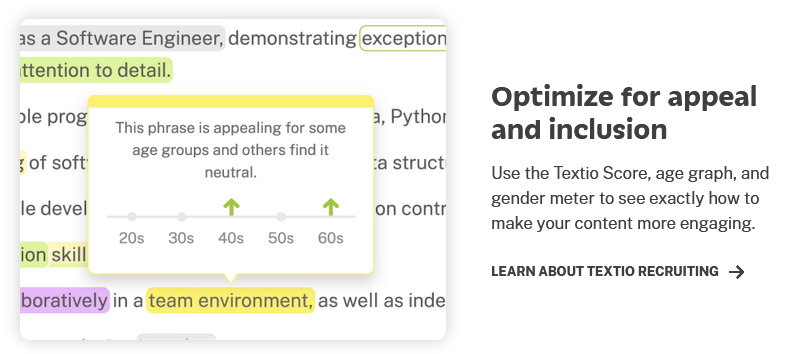}
  \captionof{figure}{A screenshot of Textio's Website.}
         \Description[A screenshot of marketing materials from the HR Tech Company, Textio. Image follows.]{A screenshot of marketing materials from the HR Tech Company, Textio. Image follows.}
  \label{fig:textioengagement}
\end{minipage}%
\begin{minipage}{.5\textwidth}
  \centering
  \includegraphics[width=.99\linewidth]{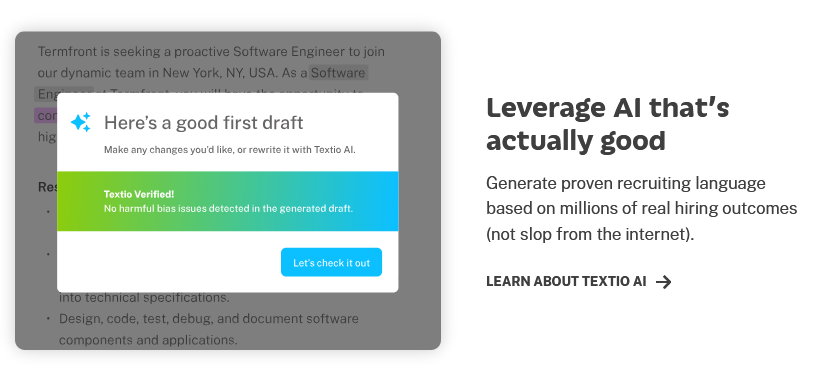}
  \captionof{figure}{A screenshot of Textio's Website.}
    \Description{A screenshot of marketing materials from the HR Tech Company, Textio. Image follows.}
  \label{fig:textioAI}
\end{minipage}
\end{figure*}

\subsection{Investigating Product Claims and Assumptions}
    We identified various claims that HR tech products made, examining descriptions of HR Tech products and inferring underlying claims. For example, many products offered process automation, claiming to reduce recruiters’ clerical workload. In this instance, the product is claiming that automation of various clerical processes (e.g., candidate outreach) makes recruiters work more efficiently. Assumptions were equally complex and required a more subject analysis. An assumption, as outlined in our related work section, is a set of tacit, background knowledge and taken-for-granted frameworks to any activity or action that are not critically reflected upon \cite{harvey2024cadaver,cook1999bridging}. For example, an assumption embedded into an automated tool that generates candidate correspondence would be that automation makes recruiters more efficient. Claims are specific to individual products; however, assumptions tend to be more general. \par

    \subsubsection{\textbf{Product Claims}}  
    Claims are product specific, but several major subgroups emerged during our analysis. The first major cluster was around accuracy and efficiency. Many HR Tech products claimed to increase recruiter efficiency or reduce time spent on non-candidate interactions. Such claims referenced features like generative AI for personalized candidate communication or automated scheduling. For example, Textio uses AI to improve job description quality and candidate communications by checking for bias (Figure \ref{fig:textioengagement}). Textio claims to ‘optimize’ text for ‘appeal and inclusion’ in the screenshot we've taken. Textio also claims its AI-generated content is high quality, not ‘slop’ (Figure \ref{fig:textioAI}). Such claims demonstrate an industry focus on efficiency and time savings, while not sacrificing quality.  \par

    A second cluster of claims we uncovered was products that claimed to be an end-to-end hiring platform, including both ATS and larger HRIS with integrated job candidate tracking. Claims promoted workflow synergies and integrations, which we also documented. We highlighted common integrations  (e.g., compatibility with a payroll or HRIS system) in the Recruiter Environment. We chose to represent these product claims in our summaries of the claims by asserting that this product was an ATS or an HRIS, and then discussing other claims about the product. \par

    Finally, many claims clustered around AI, with many claims focusing on a HR Tech product's AI capabilities (e.g., Textio, Figure \ref{fig:textioAI}). Claims emphasized AI compatibility with existing recruitment processes, and AI capabilities within recruitment workflows. An example of this is ADP's candidate relevance tool, which claims to help match information from a candidate's job materials and resume to keywords within the job description and provides a `relevancy' ranking. These claims stipulate familiarity with and understanding of AI and how AI makes various decisions as they relate to job candidates. These tools are also marketed as time savers or potentially as efficient substitutes for actual human evaluation of job candidates. For this category, we detailed how embedded AI functions, as practitioners also emphasized the need to ask vendors detailed questions. Pw5-1 commented, 

    \begin{quote}
        ``Maybe that's a caution you put on here somewhere as a general statement: for [recruiters/HR teams] to ask more questions, if they engage these vendors, to get them to truly discern or describe how they might be leveraging true AI.''
    \end{quote}
    
    This suggestion to build a tool that could challenge practitioners to `ask better questions' or `ask the right questions to know what they're really thinking' (P97, returning for the walkthrough interviews) of these HR Tech companies was one we took to heart, as it is at the root of the kind of contextual AI transparency that we wanted to build. This approach informed how we documented hidden assumptions in products. \par

\subsubsection{\textbf{Assumptions}}
    To assess the assumptions in our database, we examined product claims, functionality, and their underlying conceptions of recruiting through our understanding of the professional practices of recruiting. The assumptions we listed represent our interpretation of the assumptions each HR Tech product makes about HRM and recruiting. For example, CodeSignal, used in candidate screening, treats personality or test-taking ability as definitive indicators of candidate fit. Though these tests are commonly used, they are not sole predictors of job success. We reviewed each product website, documenting information around how products used data, as well as how they used AI, to ensure we had a complete list of assumptions. \par
    
    \begin{figure}[h]
        \centering
        \includegraphics[width=0.75\linewidth]{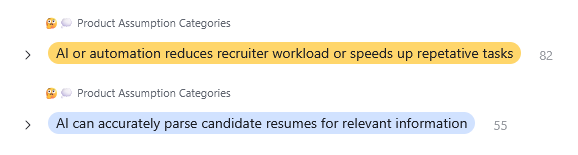}
        \caption{Screenshot of the Assumptions View of the Recruiter Environment.}
        \label{fig:Assumptions}
        \Description{A screenshot of the Assumptions View of the Recruiter Environment. Image follows.}
    \end{figure}

    After completing the companies in our database, we performed a close read of the assumptions we had identified and created 24 clusters of assumptions. For example, many products used generative AI to personalize candidate communication. The broader categories accounted for this, forming clusters such as `AI can produce more appealing written content than humans' and `humans respond well to generative AI messages.' Figure \ref{fig:Assumptions} shows how these assumptions were represented within the Researcher Environment. PW1-1 (P76), upon reviewing the Research Environment, commented: 
    
    \begin{quote}
    ``[S]ome of [the assumptions] flag me [...], like, \textbf{\textit{`AI can produce more appealing written content than humans,}}' and my knee-jerk reaction is, \textbf{\textit{`No, I don't like that.'}} And it helps to have [explainer text] this above every single [assumption category] so I remember this is an assumption; this isn't this website making this claim to me.'' \end{quote}\par

    \begin{table*}
        \centering
\begin{tabular}{|l|l|c|}
\hline
\multicolumn{1}{|c|}{\textbf{Assumptions Cluster}} & \multicolumn{1}{c|}{\textbf{Example Tags}} & \textbf{\%} \\ \hline
\begin{tabular}[c]{@{}l@{}}How humans interact with HR \\ Tech Products as either users \\ or candidates.\end{tabular} & \begin{tabular}[c]{@{}l@{}}Chatbots can appropriately and effectively \\ converse with candidates.\\ AI or automation reduces recruiter workload or \\ speeds up repetitive tasks\end{tabular} & 13.04\% \\ \hline
\begin{tabular}[c]{@{}l@{}}How humans interact with the \\ outputs of HR Tech products.\end{tabular} & \begin{tabular}[c]{@{}l@{}}AI can determine which candidates are the best \\ fit for a role based on job materials\\ Candidates respond well to AI-generated \\ messages.\end{tabular} & 17.39\% \\ \hline
\begin{tabular}[c]{@{}l@{}}How HR Tech Products and \\ their AI Systems Function\end{tabular} & \begin{tabular}[c]{@{}l@{}}AI can accurately parse candidate resumes for \\ relevant information\\ Automated communication improves recruiter \\ effectiveness\end{tabular} & 34.78\% \\ \hline
\begin{tabular}[c]{@{}l@{}}How practitioners understand \\ the outputs of AI systems \\ embedded into HR Tech \\ products.\end{tabular} & \begin{tabular}[c]{@{}l@{}}AI is less biased than humans and will act more \\ fairly in hiring contexts\\ Humans understand AI candidate ranking \\ and will evaluate that ranking critically\end{tabular} & 34.78\% \\ \hline
\end{tabular}
\caption{A table describing categories assumptions embedded into HR Tech Products }
\label{tab:AssumptionsCategories}
\Description{A table describing categories assumptions embedded into HR Tech Products. Table follows.}
    \end{table*}    

\begin{figure}
        \centering
        \includegraphics[width=0.5\linewidth]{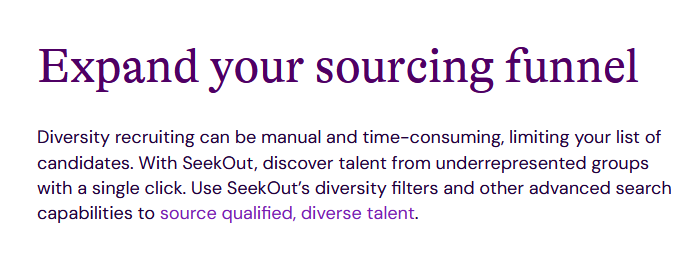}
        \caption{A screenshot of Seekout, a HR Tech product's website, advertising its technical tools to help find diverse job candidates}
        \label{fig:seekoutdiv}
        \Description{A screenshot of Seekout, a HR Tech product's website, advertising its technical tools to help find diverse job candidates. Image follows.}
    \end{figure}

\begin{figure}
        \centering
        \includegraphics[width=0.5\linewidth]{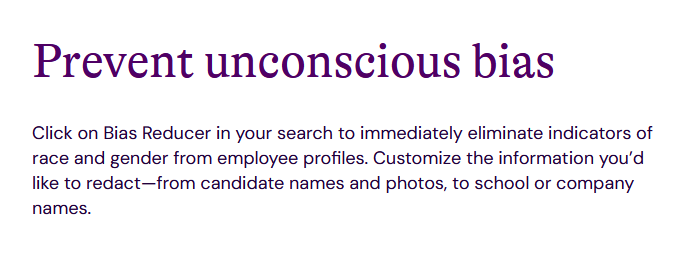}
        \caption{A screenshot of Seekout, a HR Tech product's website, advertising its technical tools to hide indicators of diversity to prevent unconscious bias.}
        \label{fig:seekouthide}
         \Description{A screenshot of Seekout, a HR Tech product's website, advertising its technical tools to hide indicators of diversity to prevent unconscious bias. Image Follows}
    \end{figure}

    We identified meta-clusters that extend beyond general assumptions about HRM among technology companies.  A list of these is seen in Table \ref{tab:AssumptionsCategories}. A major area of focus that cross-cut these clusters was companies framing bias as a human, not a technology, problem.. For example, Textio treats bias as a human input issue (see Figure \ref{fig:textioengagement}), as the product is designed to correct the `biased' writing of humans. Conversely, SeekOut treats bias as both human and technical, identifying diversity indicators overlooked by job boards and allowing concealment to prevent unconscious bias (see Figures \ref{fig:seekoutdiv}, \ref{fig:seekouthide}). Both of these products are used during the sourcing stage of hiring, with Textio overlaying other products to augment communication, and SeekOut identifying candidates that may have otherwise been overlooked. Contrasting these assumptions shows how HR Tech products shape user interaction with them and could influence policy workarounds. \par

    During the walkthrough interviews with HR management and procurement professionals, W3-4, an HR Manager, mentioned how challenging it was to find candidates who had specific requirements, while still adhering to policies in her office around handling job candidates. Other participants echoed the need for more decision-making information. The assumptions were a good place to find such information. As such, we kept assumptions as both a long-form text field and a tagged system, for comparison, Tagging enabled filtered claims views; long-form text captured nuanced categories. \par

\subsection{Assessing AI Clarity}
    Understanding product functionality, claims, and assumptions gave us the foundation to assess the clarity of a product's AI. We determined this through two assessments: whether the AI's purpose within the tool was clearly described, and whether the AI's function was clearly described. We reviewed product marketing to see if the AI’s presence and purpose was clear to an average user. For example, we encountered many HR Tech products that seemed to simply say they had AI without clear justification or proof of its presence. As P100 pointed out, `\textit{`Is it really AI? All these companies are saying AI this, AI that; a lot of it is just automation.''} Similarly, PW5-1 pointed out, \textit{``[HR Tech products] sometimes will reference automation as AI.''} Secondly, after documenting product functionality and claims, we had a clear picture of how the HR Tech products we examined conceptualized AI and how transparently their marketing materials described its function. Our second measure was whether we could discern actual AI functionality from the existing marketing materials. \par

    As we investigated the products, we used the following decision process. If the AI embedded into a product's purpose and functionality was clear to the research team, we marked the AI clarity as `clear.' If we encountered conflicting information or vague descriptions (e.g., Textio's ``AI that doesn't suck'') around either purpose or functionality, we marked the product's AI as `somewhat clear.' If we couldn't determine either product or functionality, or there was not enough information available to understand the AI, we labeled the product as `unclear.' We also labeled products with no AI. After workshop feedback, we included reasoning behind each ranking. Table \ref{tab:AIDescBool} describes this decision process. 

    \begin{table*} [h]
    \begin{tabular}{|l|l|l|}
\hline
\textbf{Ranking} & \textbf{Purpose (To What End)} & \textbf{Function (How)}    \\ \hline
\rowcolor[HTML]{D9EAD3} 
Clear            & Explained                      & Explained                  \\ \hline
\rowcolor[HTML]{FFF2CC} 
Somewhat Clear   & Explained or Not Explained     & Explained or Not Explained \\ \hline
\rowcolor[HTML]{F4CCCC} 
Unclear          & Not Explained                  & Not Explained              \\ \hline
\end{tabular}
\caption{Clarity of AI Product Description Boolean}
\Description{A table detailing clarity of AI product description boolean.}
\label{tab:AIDescBool}
    \vspace{-5mm}
    \end{table*}   

    Participants found these clarity rankings useful for understanding the embedded AI within the HR Tech products we examined. PW4-1, who works as an influencer within the HR space and as a recruiter, suggested that these rankings could also be useful for HR Tech developers who want to make their marketing materials more transparent. He explained:        
        \begin{quote}
            \textit{``To be able to help [HR Tech companies] understand well, you got an Unclear because [the research team] couldn't find anything [about how your AI worked] on your website. [...] You would be helping them, from a marketing standpoint, to say, hey, look, you're not articulating these messages very well and we're talking about AI in many cases, so the explainability and defensibility around that [clarity] is incredibly important.''}
        \end{quote}
        
    Comments during the persona workshops also asked about database maintenance to ensure for long-term value to practitioners and vendors. While our end goal was to create a practice-oriented tool to help practitioners assess AI clarity, this is an added potential benefit for our work. Using our three-point scale to assess AI clarity, as we navigated through our investigation, we were able to illustrate how intransparency around AI used in HR has extended beyond the technical into spaces where such intransparency could potentially be misleading to consumers. This approach was not without its flaws, however, and it was limited in several ways, as we explore in the next section.  \par

    \subsection{Limitations to Our Approach}
    One of the challenges of building a database of technology products is that there is no set way that they are described in their marketing materials. As mentioned in our literature review, there is a relatively similar structure to the data that facilitates HR Tech functionality \cite{ajunwa2019platforms}, but, unfortunately, this data structure does not extend to marketing language. By limiting ourselves to only what we could identify from first-person (so to speak) data sources, we encountered moments where there simply was no data to enter into the database. For example, collecting basic information about the HR Tech products we evaluated was not always clear-cut. Many products had multiple functionalities, and circular promotional websites that made finding information about product functionality challenging. To address this limitation in our approach, we built out inclusive product tag lists so that products could exist in more than one category based on the hiring stage they supported. But even this design decision had its limitations, when the information needed to complete the HR Tech product's documentation was simply not there in the first place. This was most evident when investigating AI functionality, as `AI' functioned like a catch-all in many instances that did not allow for the specificity needed in our documentation. Often, when we encountered this circumstance, we would enter `unclear' into whatever detail we were investigating. We acknowledge that, by limiting our data sources to only products' websites, we were not able to provide complete and accurate information. Or, the information we received was based on misleading or unclear information on a product's company website. During the practitioner workshops, PW2-3, a HR Manager, asked us for clarification on our labeling of the product ADP:

    \begin{quote}
        ``I'm not the most technical person. I will admit that I don't know what I'm talking about, but \textbf{\textit{I use the AI feature within ADP the most out of any one of our systems}}. I use it to create job descriptions. I use it to generate responses to the applicants. I use it for parsing and filtering. \textbf{\textit{So I'm so confused when I'm looking on here and seeing no AI!}}''
    \end{quote}

    When we re-investigated what we had documented about ADP, we found that what had initially been interpreted as a third-party AI product integration available through ADP's internal add-on marketplace was in fact a native candidate ranking service. Making sense of marketing materials that were so opaque that we could not discern functionality or with marketing language shifting as new products were rolled out or companies merged, often led us to use the catchall `unclear' category, which was as opaque as the product information we were trying to understand. Further, information and marketing claims on websites can change quickly -- as each update to ad copy does not necessarily have to reflect a change in functionality, just a shift in marketing language. We took steps to mitigate this flaw in our approach by detailing \textit{why} things were unclear or opaque when we had to add an `unclear' descriptor, but acknowledge that this approach only goes so far as people's willingness to read our reasoning.\par

\subsubsection{\textbf{Drawbacks to AI Clarity Rankings}}
    Another flaw we found in our method was in how we documented AI clarity. While participants were interested in the complexities of AI clarity rankings, they also pointed out that they would probably just use those ratings to identify the `top' products that met their needs and assess those with their senior management teams. W3-2 explained, \textit{``Basically a director wants me to get the top three [HR Tech Products]. So what I do to get those top three is go through the clear [tag] and then maybe download a demo.''}. Similarly, W3-4 explained, \textit{[If we] have a catalog of companies that we are vetting, I would get a little overwhelmed with the view all list [in the database]. [...] At the end of the day, I'll probably narrow it down to the top three, top five. I know the list is long because you gathered all the lists possible, but I'm just saying, I would need to know who's best with what [I need.]''} In both of these cases, the walkthrough interview participants admitted that they would default to the top few items on the list that would meet their requirements with a `clear' ranking. The implied value of a `clear' ranking, as PW5-1 hinted at above, suggests that a `clear' AI Clarity Ranking would automatically put a product at the top of a list for procuring licenses, regardless of the embedded claims, assumptions, or other details of these products. \par

    In addition to this were instances where some aspect of a product's AI was explained clearly, but other parts were not. In these instances, we would label the product as `somewhat clear' despite there being parts of the AI's functionality that were explicitly explained and well-documented. By providing a ranking at the product level, as opposed to a more finite ranking, we limited the effectiveness of these ratings, and potentially would cause procurement teams to miss potentially useful products based on a perceived `negative' rating. \par

\section{Conclusions and Future Work}
    Our findings describe a method for building a practice-oriented database of available tools as a means by which to explore our research questions: (1) What is the Recruiting HR Tech space?, (2) How is AI deployed within these products?, and (3) What is being conveyed about AI use within these products? By focusing on basic elements of any technology product that uses AI: functionality, product claims, underlying assumptions, and overall AI clarity, we demonstrate how contextual AI transparency can be achieved through database design. We drew on interviews with practitioners to inform our prototype and evaluative workshop discussions of our prototype to create the TARAI Index using an iterative approach to situate the information we gathered to create an assessment of each HR Tech product we reviewed. \par
                 \begin{table*}[h]
        
\centering
\begin{tabular}{|l|l|}
\hline
\multicolumn{1}{|c|}{\textbf{Investigative Angle}} & \multicolumn{1}{c|}{\textbf{Intransparency Finding}} \\ \hline
\textit{Functionality} & \begin{tabular}[c]{@{}l@{}}Murky descriptions of AI, vague language and conflating terms \\ make product and AI functionality hard to comprehend.\end{tabular} \\ \hline
\textit{Claims} & \begin{tabular}[c]{@{}l@{}}Unclear connections between functionality and claims leads to \\ transparency issues\end{tabular} \\ \hline
\textit{Assumptions} & \begin{tabular}[c]{@{}l@{}}AI is assumed to function in various ways, which can create \\ transparency issues\end{tabular} \\ \hline
\textit{AI Clarity} & \begin{tabular}[c]{@{}l@{}}When descriptions of product functionality, claims, or \\ underlying assumptions are vague, AI is also described in \\ intransparent ways.\end{tabular} \\ \hline
\end{tabular}
\caption{A table describing our findings of how intransparency is expressed.}
\Description{A table describing our findings of how intransparency is expressed. Table follows.}
\label{tab:intransparency}

    \end{table*}

    We grounded this research in the social practice of recruiting, engaging practitioners to understand tensions between the materials, competencies, and meanings of their professional practices and the HR Tech products they used. Our understanding of recruiting allowed us to see the nuances between product functionality and product claims, as practitioners repeatedly noted discrepancies in this regard. This understanding also helped us to assess the underlying assumptions of these products, where we found that many products' assumptions often assign responsibility for bias to recruiters, rather than tech or design flaws. In sum our findings reveal ongoing challenges for recruiters due to opaque AI systems. By situating ourselves within the social practice of recruiting, we were able to articulate and create a tool for contextual AI transparency.

    Unlike prior transparency databases focused on failures or harmful incidents \cite{mcgregor2021AIIncident,walker2024deepfakedb}, our approach offers a cross-cutting view of AI within recruiting as means of creating social and developer awareness, our approach is novel in that it allows for a cross-cutting view of the AI used within a specific professional practice - recruiting within the broader field of HRM. By examining product functionality, claims, underlying assumptions and AI clarity, we can provide practitioners with a usable tool for knowledge building and meaning-making within their HR Tech product stack, and for researchers, we can demonstrate a concerning pattern we have seen in AI interventions in Tech products: that AI in transparency isn't just a technical black box, but rather a set of choices around language and assumptions about people as well. \par

    The method we used to build the TARAI Index allowed us to explore transparency within the HR Tech space, and our investigation of what these products say about themselves revealed ongoing issues of opacity within the HR Tech space. With each of the four major areas that we investigated: functionality, claims, assumptions, and AI clarity, we found countless moments where marketing materials did not accurately reflect how a product worked, claimed things that were unprovable with their existing marketing copy, and were vague about what data was used to train its AI, or how that AI functioned. Table \ref{tab:intransparency} reflects our observations, summarized. Many of the HR Tech products lock full functionality documentation and more information about their AI behind connections with a sales team for demonstrations, or by providing an email address or phone number. For the HR Tech products that did make such information available, it was often buried in privacy policies or hard to locate documentation. \textbf{We urge designers to take note that walling off product information that could lead to better AI transparency by marketing teams is the norm, rather than the exception. We ask that designers reflect on what is gained by placing important product information, such as a full list of functionalities to back up marketing claims, behind walls managed by gatekeepers who wish to sell a product, and who may only be willing to speak to people with purchasing power within an organization.} Contextual AI transparency in HR Tech requires openly accessible product details, not gatekept conversation. 

    We welcome and encourage future work that explores this approach to creating contextual AI transparency within various professional practice settings. For example, with diagnostic AI coming onto the market from developers such as Epic \cite{ICTHealth2025EpicAI} for use in hospitals, understanding how these tools were built and on what assumptions they were built upon is vital. Other applications could extend to the legal or creative feilds. We believe this method allows for more cross-cutting observations of a field and can function almost as a health check of the integrated AI.
    
    In this paper,  we described a practitioner-oriented technology database designed to address questions AI integration into HR Tech. We explored HR Tech product functionality, claims, assumptions, and clarity in a database through iterative design Multiple evaluations revealed that AI intransparency is both a technical and a social issue. Our contributions include these insights into HR Tech specifically and a novel approach for contextual AI transparency. We hope this method will take hold in creating similar contextual AI transparency databases for other professions. \par

\begin{acks}
[Removed for Anonymous Review]
%%Michael, Sarah, Roshni, Sheilla, Ceilia, Darden, 
\end{acks}

\bibliographystyle{ACM-Reference-Format}
\bibliography{00References}

%\clearpage

 \clearpage

\appendix
\section{Appendix A: Evaluation Methods and Participant Tables}
\begin{table*} [h]
\begin{tabular}{|l|c|}
\hline
\multicolumn{1}{|c|}{\textbf{Persona}} & \multicolumn{1}{c|}{\textbf{Persona Details}}                                                                                                                                                                                                                                                                                                                                                                      \\ \hline
Alex the AI Auditor                    & \begin{tabular}[c]{@{}l@{}}$\bullet$ Clarity on specific functionality of AI models, including \\ clear information on training data and data used in operation.\\  $\bullet$ Archive of existing audits that have taken place on this \\technology.\\  $\bullet$ Up-to-date list of applicable local, state, and federal AI \\regulations that may apply to this product.\end{tabular}                             \\\hline
Reese the Recruiter                    & \begin{tabular}[c]{@{}l@{}}$\bullet$ Clear statements of what this product does, and existing \\ integrations with other technologies such as Human Resources \\Management Suites or  Applicant Tracking Systems.\\ $\bullet$ Ease of Use. \end{tabular}                                                                                                                               \\ \hline
\begin{tabular}[c]{@{}l@{}} Peyton the HR \\Procurement Professional\end{tabular} & \begin{tabular}[c]{@{}l@{}}
$\bullet$ Up-to-date list of applicable local, state, and federal AI \\ Regulations that may apply to this product for internal \\risk assessment. \\ $\bullet$ Clear statements of what this product does, and existing \\integrations with other technologies such as Human Resources \\ Management Suites or Applicant Tracking Systems.\end{tabular} \\ \hline
\end{tabular}
\caption{Workshop 1 Personas}
\Description{A table detailing three personas of HR Tech users, as well as their needs from a database of HR Technologies.}
\label{tab:Personas}
\end{table*}

    \begin{table}[h]
    \begin{tabular}{|c|c|l|}
\hline
\textbf{P\#} & {\textbf{Workshop Session}} & \textbf{Participant \& Job Role}                                                                                                                         \\ \hline PW1-1 (P76) & 
1                                               & PW1-1. Senior Technical Recruiter                                                                                                                           \\ \hline  \begin{tabular}[c]{@{}l@{}}PW2-1 \\ PW2-2 \\ PW2-3 \\ PW2-4\end{tabular} &
2                                               & \begin{tabular}[c]{@{}l@{}}PW2-1. Managing Director\\ PW2-2. Senior Campus Recruiter\\ PW2-3. Recruitment Manager\\ PW2-4. Talent Acquisition Recruiter\end{tabular} \\ \hline PW3-1 &
3                                               & PW3-1. Technology Developer                                                                                                                                 \\ \hline PW4-1 &
4                                               & PW4-1. HR Influencer                                                                                                                                        \\ \hline PW5-1 &
5                                               & PW5-1. Chief Risk \& Compliance Officer                                                                                                                     \\ \hline
\end{tabular}
\caption{Participant job roles for the practitioner workshops.}
\Description{A table detailing participant job roles for the practitioner workshops.}
\label{tab:Workshop2Participants}
    \vspace{-5mm}
    \end{table}

    \begin{table}[h]
    
\begin{tabular}{|c|c|c|c|}
\hline
\textbf{P\#} & \textbf{Job Role}                                                                         & \textbf{Age} & \textbf{Reported Gender} \\ \hline
W3-1         & \begin{tabular}[c]{@{}c@{}}{[}Anonymized HRIS{]}\\ Operations Manager\end{tabular}        & 62           & Woman                    \\ \hline
W3-2         & HR Director                                                                               & 54           & Man                      \\ \hline
W3-3         & Early Careers Leader                                                                      & 53           & Woman                    \\ \hline
W3-4         & Human Resources Director                                                                       & 55           & Woman                    \\ \hline
W3-5         & VP Human Resources   & 55           & Woman                    \\ \hline
\end{tabular}
\caption{Participant job roles for the interviews with Senior HR Professionals in Workshop 3.}
\Description{A table detailing participant job roles for the elite interviews conducted in Workshop 3.}
\label{tab:Workshop3Participants}
    \vspace{-5mm}
    \end{table}

\end{document}